\documentclass[10pt,twocolumn]{article}
\usepackage[top=1in, bottom=1in, left=1in, right=1in]{geometry}
\usepackage{palatino}
\usepackage{amsmath, amsthm, amssymb}
\usepackage{graphicx}
\usepackage{epstopdf}
\usepackage{overpic}
\usepackage{cancel}
\usepackage{rotating}
\usepackage{url}
\usepackage{caption}
\usepackage{color}
\usepackage{rotating}
\usepackage{multirow}
\usepackage{wrapfig}

\usepackage[bottom,flushmargin,hang,multiple]{footmisc}
\usepackage{lipsum}

\newcommand{\bx}{\mathbf{x}}
\newcommand{\dt}{\Delta t}
\newcommand{\bX}{\mathbf{X}}
\newcommand{\bY}{\mathbf{X'}}
\newcommand{\bA}{\mathbf{A}}
\newcommand{\bAtilde}{\mathbf{\tilde{A}}}
\newcommand{\bU}{\mathbf{U}}
\newcommand{\bSigma}{\mathbf{\Sigma}}
\newcommand{\bV}{\mathbf{V}}
\newcommand{\bW}{\mathbf{W}}
\newcommand{\bLambda}{\mathbf{\Lambda}}
\newcommand{\bPhi}{\mathbf{\Phi}}
\newcommand{\imag}{\text{imag}}
\newcommand{\bL}{\mathbf{L}}
\newcommand{\ba}{\mathbf{a}}
\newcommand{\bz}{\mathbf{z}}

\definecolor{blue}{rgb}{0,0,1}

\graphicspath{{./}}

\title{Extracting spatial-temporal coherent patterns in large-scale neural recordings using dynamic mode decomposition}
\author{Bingni W. Brunton$^{1,2}$\thanks{Contact: bbrunton@uw.edu}, Lise A. Johnson$^{3,4}$, Jeffrey G. Ojemann$^4$, J. Nathan Kutz$^2$\\
\\
\small{$^1$ Department of Biology, $^2$ Department of Applied Mathematics, $^3$ Center for Sensorimotor Neural Engineering,} \\ 
\small{$^4$ Department of Neurological Surgery, University of Washington, Seattle WA 98195, USA}}
\date{}
\begin{document}

\maketitle

\begin{abstract}
There is a broad need in the neuroscience community to understand and visualize large-scale recordings of neural activity, big data acquired by tens or hundreds of electrodes simultaneously recording dynamic brain activity over minutes to hours.
Such dynamic datasets are characterized by coherent patterns across both space and time, yet existing computational methods are typically restricted to analysis either in space or in time separately.
Here we report the adaptation of dynamic mode decomposition (DMD), an algorithm originally developed for the study of fluid physics, to large-scale neuronal recordings.
DMD is a modal decomposition algorithm that describes high-dimensional dynamic data using coupled spatial-temporal modes; the resulting analysis combines key features of performing principal components analysis (PCA) in space and power spectral analysis in time.
The algorithm scales easily to very large numbers of simultaneously acquired measurements.
We validated the DMD approach on sub-dural electrode array recordings from human subjects performing a known motor activation task.
Next, we leveraged DMD in combination with machine learning to develop a novel method to extract sleep spindle networks from the same subjects.
We suggest that DMD is generally applicable as a powerful method in the analysis and understanding of large-scale recordings of neural activity.
\end{abstract}

%

\section{Introduction}\label{sec:Intro}

Advances in technology and infrastructure are delivering the capacity to record signals from brain cells in much greater numbers and at even faster speeds. 
This deluge of data is central to answering many critical open questions in neuroscience and motivates the continued development of computational approaches to analyze, visualize, and understand large-scale recordings of neural activity.
Fortunately, the activity of complex networks of neurons can often be described by relatively few distinct patterns (for instance,~\cite{Broome:2006, Yu:2009, Machens:2010, Churchland:2012, Shlizerman:2014}).
Identifying these spatial-temporal patterns enables the reduction of complex measurements through projection onto coherent structures, where it is tractable to build dynamical models and apply machine learning tools for pattern analysis.
Here we introduce dynamic mode decomposition (DMD) as a novel approach to explore spatial-temporal patterns in large-scale neural recordings.
The method combines well-characterized advantages from two of the most powerful data analytic tools in use today: power spectral analysis in time and principal components analysis (PCA) in space.

Measurements of neural activity from tens to hundreds of simultaneously recorded channels are traces in time that probe a network with complex dynamics, and one principled way to make sense of such dynamic networks is with modal decomposition~\cite{Holmes:Dynamics}.
A particularly popular modal decomposition tool is PCA, which derives modes ordered by their ability to account for energy or variance in the data~\cite{Jolliffe:PCA}.
PCA has already been widely applied in the study of high-dimensional biological systems; 
however, it suffers from a few well known drawbacks.
In particular, PCA is a static technique and does not model temporal dynamics of time-series data explicitly, so it often performs poorly in reproducing dynamic data, such as recordings of neural activity.

Neural dynamics are well known to be characterized by dynamic oscillations at many frequency bands, which are implicated in a variety of neural functions~\cite{Raghavachari:2001, Buzsaki:2004, Fries:2005, Uhlhass:2010}.
Most tools analyzing the frequency content of a signal are related to the Fourier transform, which 
transforms time-varying signals into a spectrum in the frequency domain.
Importantly, the power spectrum can be computed  efficiently using the fast Fourier transform (FFT) algorithm~\cite{Welch:1967}, whose efficient implementation has contributed to its ubiquitous use.
One example of a modal decomposition in time that goes beyond the Fourier transform is  empirical mode decomposition (EMD), which computes intrinsic oscillatory modes from time-varying data~\cite{Huang:1998}.
EMD has been used to analyze neural data, including cortical local field potential~\cite{Liang:2005} and EEG~\cite{Sweeney:2007}.
However, these time-domain transforms are typically applied to individual signals (see exceptions in \cite{Rehman:2009}) and do not support spatial structures.

A relatively new modal decomposition method is DMD~\cite{Rowley:2009,Schmid:2010}.
DMD was developed initially to study experiments and simulations in the fluid mechanics community, where it was introduced to reduce very high-dimensional dynamic data into relatively few coupled spatial-temporal modes.
Importantly, it has been shown that DMD is related to Koopman spectral analysis, motivating its usefulness in characterizing dynamics of nonlinear systems~\cite{Rowley:2009}.
More recently, DMD has been generalized as an eigendecomposition of an approximating linear operator~\cite{Tu:2013}, allowing it to be more easily applied to a larger class of datasets.

To demonstrate DMD's applicability to large-scale neural recordings, we analyzed sub-dural electrode array recordings from human subjects in two different contexts.
First, we validated the DMD approach to derive sensorimotor maps based on a simple movement task~\cite{Miller:2007}.
Next, we leveraged DMD in combination with machine learning techniques to detect and characterize spindle networks present during sleep; a method to automatically extract these stereotyped sleep spindle networks had not been described previously in the literature.

We show that sleep spindles tended to occur coincidentally in different groups of electrodes at different times~\cite{Johnson:2012}.
These patterns may reflect some underlying anatomical or functional connectivity in the brain, and studying relatively local spindle networks may be a novel way to understand the organization of the brain.
Our spindle network detection algorithm enables the exploration of sleep spindle networks in a large number of experimental subjects, which will shed light on thalamacortical connections and local cortical networks.

The purpose of this work is to describe DMD and its adaptation to the analysis of large-scale neural recordings.
For very large datasets whose dimensionality strains typical computing resources, DMD may be readily implemented using routine linear algebra functions to take advantage of cluster computing~\cite{Freeman:2014}.
We suggest that DMD may be useful in understanding spatial-temporal coherent patterns in data of escalating scale in neuroscience, including non-invasive and invasive measurements such as functional MRI, MEG, neurophysiological recordings with electrode arrays, and optical imaging of neural activity.


\section{Results}



\subsection{Dynamic mode decomposition (DMD)} \label{ssec:exactdmd}

Here we briefly summarize the DMD algorithm~\cite{Rowley:2009, Schmid:2010, Tu:2013} and then describe a few adaptations useful for its application to large-scale recordings of neural activity.

Consider measurements taken from $n$ observable locations at times $k \dt$, where we arrange measurements at snapshot $k$ to make a column vector $\bx_k$.
For instance, these measurements may be voltages from $n$ channels of an electrode array sampled every $\dt$.

Gathering measurements from $m$ points in time, we may construct two $n \times (m-1)$ raw data matrices:
\begin{align}
\bX &= \begin{bmatrix}
\vline & \vline & & \vline \\
\bx_1 & \bx_2 & \cdots & \bx_{m-1}\\
\vline & \vline & & \vline
\end{bmatrix}, \nonumber \\
\bY &= \begin{bmatrix}
\vline & \vline & & \vline \\
\bx_2 & \bx_3 & \cdots & \bx_{m}\\
\vline & \vline & & \vline
\end{bmatrix}.
\label{eq:datamatrices}
\end{align}
Note that $\bX$ and $\bY$ contain largely overlapping data, differing in that columns of $\bY$ are shifted one $\dt$ from those in $\bX$.

Let us suppose that there is an unknown linear operator $\bA$ such that
\begin{align}
\bY = \bA \bX.
\label{eq:yax}
\end{align}
The \emph{dynamic mode decomposition} of the data matrix pair $\bX$ and $\bY$ is given by the eigendecomposition of $\bA$.
We may think of $\bA$ as describing a high-dimensional linear regression of the nonlinear dynamics which relate $\bX$ to $\bY$.

To obtain an approximation of $\bA$, one approach is to use the singular value decomposition (SVD) of the data matrix $\bX= \bU \bSigma \bV^*$ to compute its pseudoinverse:
\begin{align}
\bA \approx \bY \bX^{\dagger} \triangleq \bY \bV \bSigma^{-1} \bU^*.
\end{align}
However, if $n$ is large, computing the eigendecomposition of the $n \times n$ matrix $\bA$ may be prohibitively expensive.
Instead, the following procedure allows the DMD modes and eigenvalues to be calculated without direct computation of $\bA$.

\vspace{10pt}
\noindent \hrulefill

{\bf DMD Algorithm} \cite{Tu:2013}
\begin{enumerate}
\item Compute the SVD of our first data matrix, 
$\bX = \bU \bSigma \bV^*$.
We may now make the substitution into Equation \eqref{eq:yax} and write $\bY = \bA \bU \bSigma \bV^*$.

\item Define $\bAtilde \triangleq \bU^* \bA \bU = \bU^* \bY \bV \bSigma^{-1}$.

\item Compute the eigendecomposition of $\bAtilde$,
\begin{align*}
\bAtilde \bW = \bW \bLambda,
\end{align*}
where $\bW$ is the matrix of eigenvectors, and $\bLambda$ is the diagonal matrix of eigenvalues.
Each eigenvalue $\lambda_i$ is a DMD eigenvalue.

\item Compute the DMD modes,
\begin{align}
\bPhi \triangleq \bY \bV \bSigma^{-1} \bW.
\end{align}
Each column of $\bPhi$ is a DMD mode $\phi_i$ corresponding to eigenvalue $\lambda_i$.
\end{enumerate}~
\noindent \hrulefill

\begin{figure*}[tb]
\centering
\includegraphics[width=0.95\textwidth]{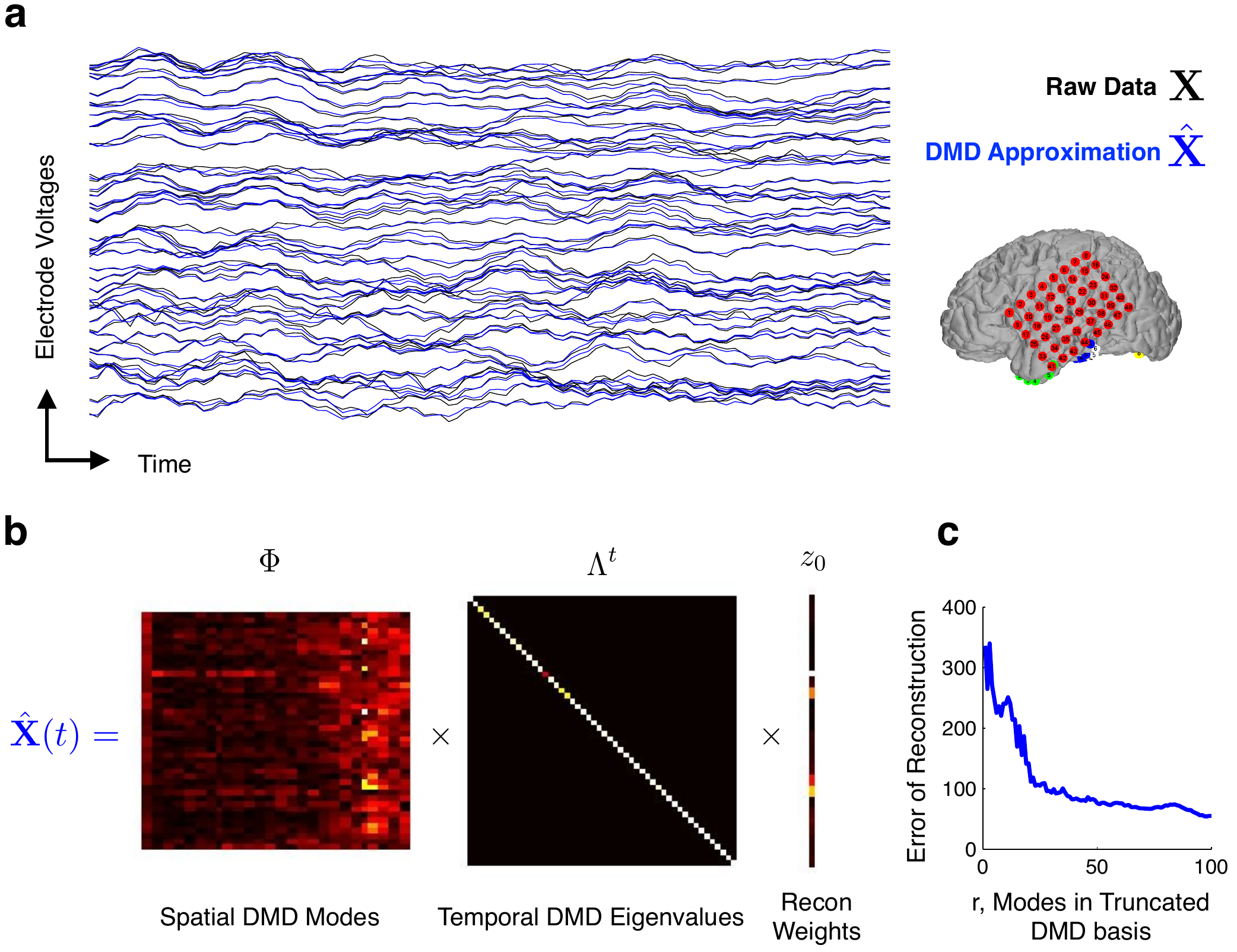}
\caption{An illustration of coupled spatial-temporal modal decomposition by DMD applied to a 500 msec window of 45-channel ECoG data sampled at 200 Hz. The raw data $\bX$ (black traces in \textbf{a}) is approximated as $\hat{\bX}(t)$ (blue traces in \textbf{a}). Voltages are shown in normalized units. \textbf{b} shows that $\hat{\bX}$ is the product of spatial modes (columns of $\bPhi$) and temporal eigenvalues $\bLambda^t$. Note that $\bPhi$, $\bLambda$, and $\bz_0$ are all complex valued, and only their absolute values are visualized in the illustration. \textbf{c} shows the decrease in reconstruction error as more modes are included in a truncated, low-rank DMD basis. The error was computed as $\|\bX - \hat{\bX}\|_F$, the Frobenius norm of the difference between the raw data and the DMD approximation.}
\label{fig:dmd}
\end{figure*}
\vspace{10pt}

\begin{figure}[h]
\centering
\includegraphics[width=0.45\textwidth]{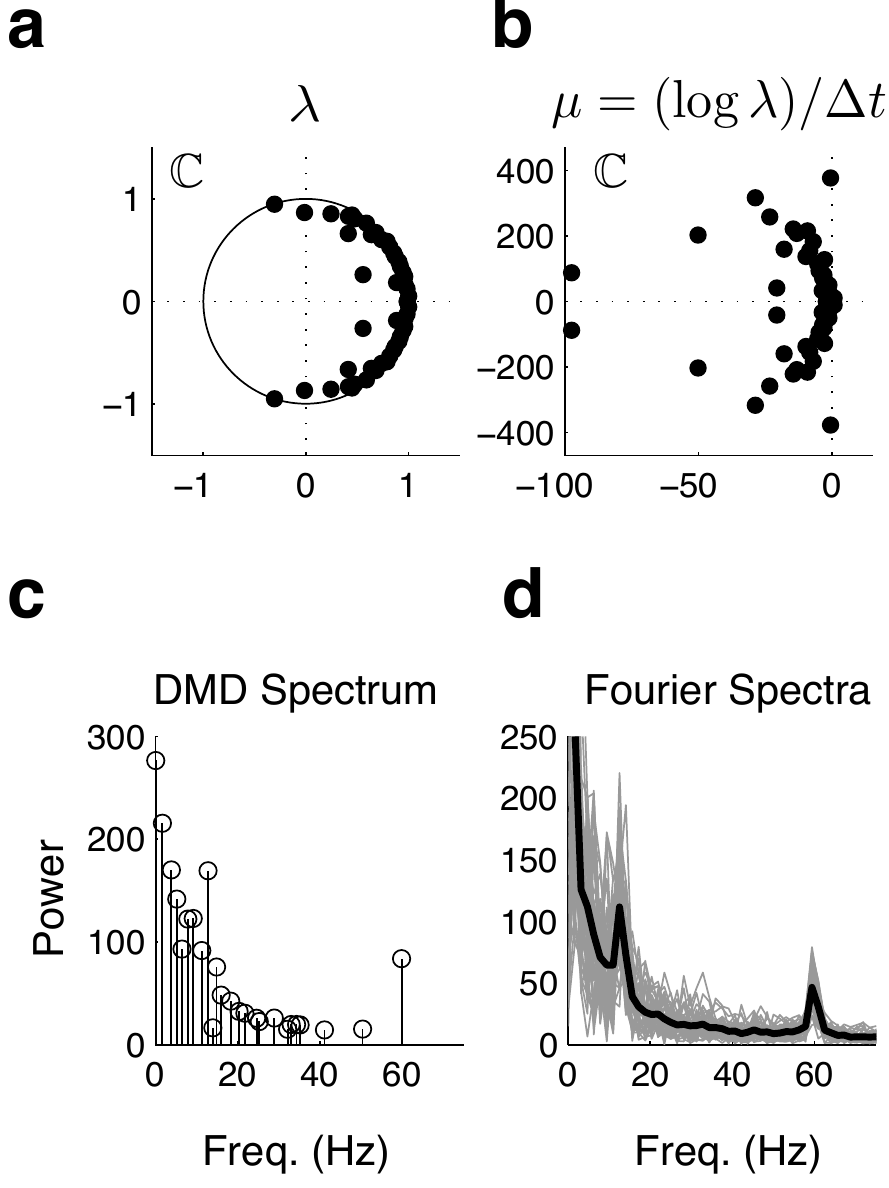}
\caption{An example of the DMD of ECoG data.
(\textbf{a}) The eigenvalues $\lambda$ visualized on the complex plane, relative to the unit circle.
(\textbf{b}) An alternate visualization of the eigenvalues, where the imaginary component of $\mu$ is proportional to the frequency of oscillation in cycles/sec.
(\textbf{c}) The DMD power spectrum, which qualitatively resembles the FFT power spectrum in (\textbf{d}). The FFT spectrum was computed independent for every recording channel (gray) and its average is shown in black. Crucially, unlike the FFT spectrum, each point in the DMD spectrum corresponds to a specific spatial mode.}
\label{fig:dmdspectrum}
\end{figure}

Finally, we may write an approximation of the observed data as a simple dynamic model $\hat{\bX}(t)$,
\begin{align}
\hat{\bX}(t) = \bPhi \bLambda^t \bz_0 ,
\label{eq:dmdrecon}
\end{align} 
where $\bz_0$ is set of weights such that $\bx_1= \bPhi \bz_0$ and can be solved as a pseudoinverse problem. 
\textbf{Figure~\ref{fig:dmd}} illustrates the separation of spatial modes $\bPhi$ from their temporal dynamics $\bLambda$ in the reconstructed multi-channel time series electrocorticography (ECoG) data.
A DMD mode $\phi_i$ is a vector that has the same dimension as $\bx$; its magnitude represents spatial correlations between the $n$ observable locations.
The eigenvalue $\lambda_i$ corresponds to the temporal dynamics of the spatial mode $\phi_i$.
Specifically, its rate of growth/decay and frequency of oscillation are reflected in the magnitude and phase components of $\lambda_i$, respectively (\textbf{Fig.~\ref{fig:dmdspectrum}a,~b}). 

The key feature of the above algorithm is that it decomposes  data, arranged as in Equation \eqref{eq:datamatrices}, into a set of \emph{coupled spatial-temporal modes}.
Note that $\hat{\bX}$ has, in general, non-zero imaginary components; if the raw data $\bX$ is strictly real valued, then we may consider only the real component of $\hat{\bX}$.

A few general properties of DMD are interesting to note. 
The data $\bx$ may be real or complex valued; in the case of recordings from electrode arrays, we will proceed assuming $\bx$ are real valued voltage measurements.
Further, the decomposition is unique~\cite{Chen:2012}, and it is also possible to compute the DMD of non-uniformly sampled data~\cite{Tu:2013}.

It is convenient to think of this spatial-temporal decomposition as a hybrid of static mode extraction by principal components analysis (PCA) in the spatial domain and spectral transformation in the frequency domain (Fourier transform).

\subsection{Low-rank DMD}\label{subject:dmdtruncated}

When the high-dimensional dynamics of the data has some underlying low-dimensional structure, 
it may be possible to capture the key dynamics of the data with relatively few DMD modes.
DMD modes are computed in the basis of $\bU$ by an eigendecomposition of $\bAtilde \triangleq \bU^* \bA \bU$, where $\bU$ is taken from a SVD of the data matrix $\bX$.
It follows that the DMD algorithm obtains exactly as many DMD modes as the number of singular values in this SVD, within numerical precision.

To obtain a truncated DMD with $r$ modes, it is possible to use instead a truncated SVD basis, keeping only the $r$ largest singular values.
Let $\bU_r$ be the first $r$ columns of $\bU$, let $\bSigma_r$ be the upper left $r \times r$ matrix of $\bSigma$, and let $\bV_r$ be the first $r$ columns of $\bV$.
This truncated SVD basis can be used in place of the full versions in steps 2--4 in the DMD algorithm; $\tilde{\bA}$ is now a $r \times r$ matrix.
Note that these low-rank DMD modes and eigenvalues are not a subset of their standard DMD counterparts, and in some cases, they can be quite different from each other.

The estimation of a truncation dimensionality $r$ to best capture signal and reject noise in a dataset is the subjective of an extensive literature (for instance,~\cite{Wold:1987, Gavish:2013}).
A simplistic heuristic is to look at the singular values on the diagonal of $\bSigma$.
We picked $r$ such that the sum of the first $r$ singular values accounts for over 95\% of the variance in the data; for 300 msec windows from the human ECoG records of around 100 electrodes sampled at 200 samples/sec, we chose $r = 40$ modes. \textbf{Figure~\ref{fig:dmd}c} shows the error of the reconstruction decreases as $r$ increases, and an examination of where this improvement plateaus also helps inform the appropriate truncation.

\subsection{Implementing DMD with an augmented data matrix}\label{subsec:dmdaugmented}

The application of the DMD algorithm to arrays of large-scale recordings of neuronal activity may require an adaptation to the normal procedure.
DMD was originally used in the study of very large fluid flow fields, where typically $n \gg m-1$.
In contrast, in neuroscience we are often interested in electrode arrays that have tens of channels sampled at hundreds of samples per second, so for a data matrix $\bX$ over a one-second window of data, $n < m-1$.
The economy SVD of $\bX$ produces $v$ non-zero singular values, where $v$ is the smaller of $n$ and $m-1$.
This property restricts the maximum number of DMD modes and eigenvalues to $n$, which is often too few to fully capture the dynamics over $m-1$ snapshots in time.

The solution to this rank mismatch is to construct modified versions of the data matrices, appending to the snapshot measurements with time-shifted versions of themselves, thus augmenting the number of channels to be $h n$.
The augmented data matrices is inspired by the Hankel matrix constructed in the eigenvalue realization algorithm (ERA)~\cite{Tu:2013}.

Specifically, we construct a new augmented data matrix $\bX_\text{aug}$, 
\begin{align}
\bX_\text{aug} &= \begin{bmatrix}
\vline & \vline & & \vline \\
\bx_1 & \bx_2 & \cdots & \bx_{m-h}\\
\vline & \vline & & \vline \\
\vspace{-5pt} \\
\vline & \vline & & \vline \\
\bx_2 & \bx_3 & \cdots & \bx_{m-h+1}\\
\vline & \vline & & \vline \\
& & \vdots  &\\
\vline & \vline & & \vline \\
\bx_h & \bx_{h+1} & \cdots & \bx_{m-1}\\
\vline & \vline & & \vline \\
\end{bmatrix}; \nonumber
\label{eq:augmenteddata}
\end{align}
and similarly for $\bY_\text{aug}$.
DMD is then applied as described above using $\bX_\text{aug}$ and $\bY_\text{aug}$ instead of $\bX$ and $\bY$.


For our multichannel recordings, we chose the smallest integer $h$ so that $h n > 2M$; this degree of stacking provided enough DMD modes to capture the observed dynamics.
The resultant augmented DMD modes now vectors of $hn$ elements; generally, the first $n$ elements of each mode are used in subsequent analyses.

\subsection{The DMD spectrum}\label{subsec:dmdspecrum}

The distribution of DMD eigenvalues $\lambda$ may be visualized relative to the unit circle on the complex plane, as in \textbf{Figure~\ref{fig:dmdspectrum}a}.
As we can see from Equation~\eqref{eq:dmdrecon}, the magnitude of the eigenvalues relative to the unit circle indicates whether the corresponding mode is growing or decaying. 
Eigenvalues exactly on the unit circle are stable modes.
The phase of each eigenvalue translates to the frequency of oscillation.
Specifically, 
\begin{align*}
f_i = \frac{\imag(\log(\lambda_i)/\Delta t)}{2\pi}
\end{align*} 
gives the frequency of oscillation of mode $\phi_i$ in units of cycles per second (\textbf{Figure~\ref{fig:dmdspectrum}c}).

The definition of DMD as the eigendecomposition of $\bA$ in Equation~\eqref{eq:yax} allows for arbitrary scaling of the DMD modes.
The algorithm outlined above  produces modes with unit norm.
To use the magnitude of these modes in a way similar to the power spectrum for selecting modes with high energy, we must choose an appropriate scaling.

One useful scaling is to multiply each mode by the corresponding singular value from the SVD.
We modified step 3 of the DMD algorithm to scale the modes by energy as follows.

\begin{samepage}
\noindent \hrulefill
\begin{enumerate}
\item[3$\dagger$.] Compute the eigendecomposition of $\hat{\bA} = \bSigma^{-1/2} \bAtilde \bSigma^{1/2}$,
\begin{align}
\hat{\bA} \hat{\bW} = \hat{\bW} \bLambda, \nonumber
\end{align}
where $\hat{\bW}$ is the matrix of eigenvectors and $\bLambda$ is the diagonal matrix of eigenvalues. The eigenvalues of $\hat{\bA}$ are identical to the eigenvalues of $\bAtilde$, and
\begin{align}
\bW = \bSigma^{1/2} \hat{\bW} \nonumber
\end{align}
are eigenvectors of $\bAtilde$ that we will use in step 4. These eigenvectors are scaled by $\bSigma^{1/2}$ so do not have unit norm.
\end{enumerate}
\noindent \hrulefill 
\end{samepage}
\vspace{10pt}

To visualize the DMD spectrum, we plot the ``power'' of each mode $\phi$ against its frequency of oscillation $f$.
The power of $\phi$ is defined as the square of its vector magnitude:
\begin{align}
P = ||\phi||^2.
\end{align}

\textbf{Figure~\ref{fig:dmdspectrum}c} illustrates the intuition that the DMD spectrum qualitatively resembles the power spectrum computed by the FFT for a 0.5 sec window of ECoG data, shown for comparison in \textbf{Figure~\ref{fig:dmdspectrum}d}.
Despite this resemblance, it is important to remember that the power spectrum is computed independently for every channel of the recording, whereas every point in the DMD spectrum corresponds to a specific spatial mode across all channels.

%
%

\begin{figure*}[t]
\centering
\includegraphics[width=0.9\textwidth]{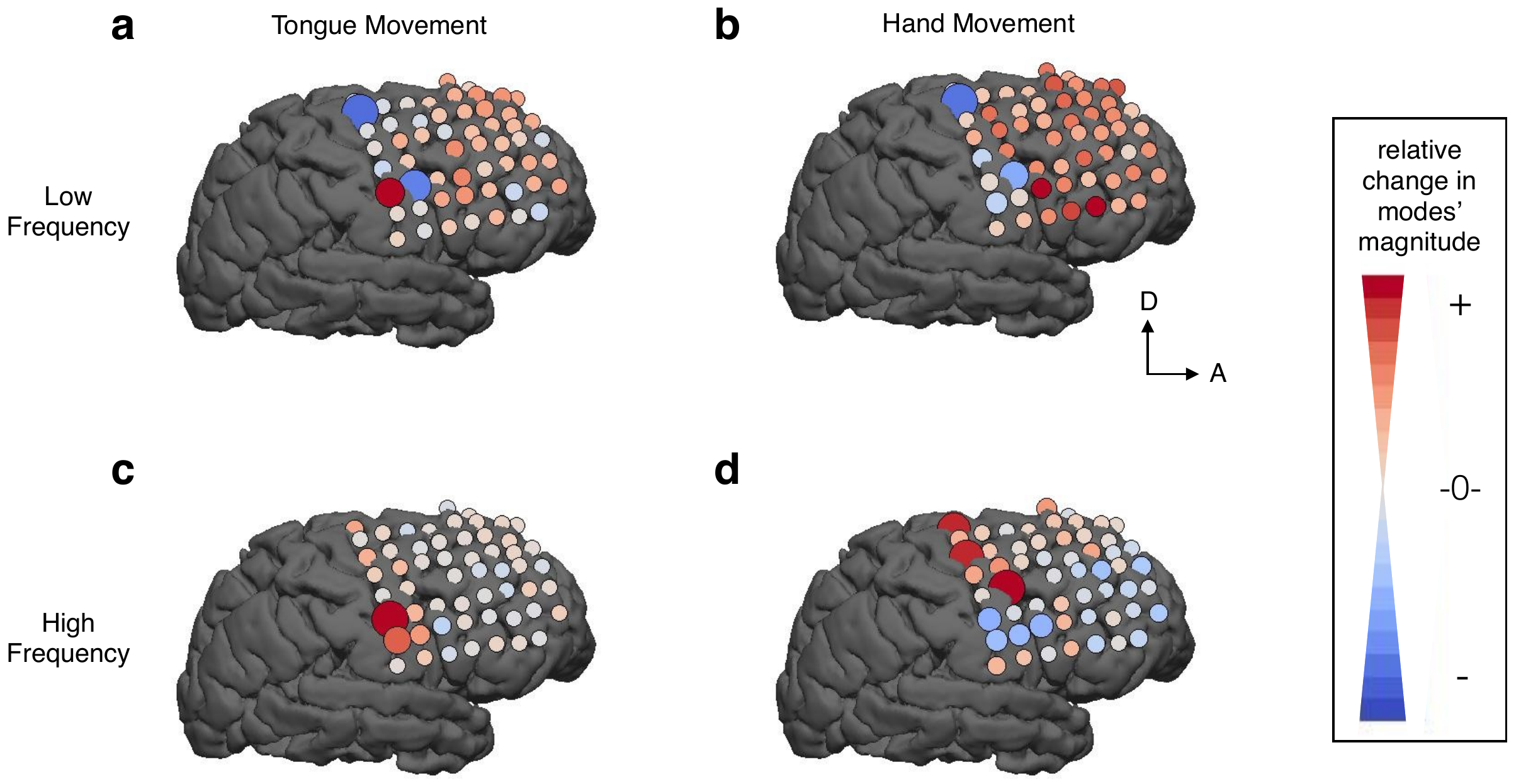}
\caption{Sensorimotor maps from Subject A, as derived from DMD modes. Relative to baseline, during movement, sites over sensorimotor cortex (the most posterior columns of the electrode array) showed decreased DMD mode magnitudes in the low frequency range (8--32 Hz, \textbf{a} and \textbf{b}), and increased magnitudes in the high frequency range (76--100 Hz, \textbf{c} and \textbf{d}). Consistent with previous literature, the sites of increased high frequency power were separable for tongue (\textbf{c}) and hand (\textbf{d}) movements.}
\label{fig:motor_validation}
\end{figure*}

\subsection{Sensorimotor mapping} \label{ssec:motor}

To demonstrate the application of DMD to large-scale time-series data, we analyzed datasets recorded from sub-dural electrode arrays implanted in two patients monitored for seizure loci localization.
To validate the DMD approach to spatial-temporal pattern analysis, we derived sensorimotor maps based on a simple motor repetition task where the subjects were instructed to move either their tongue or the hand contralateral to the implanted electrode array. 
This task was previously analyzed for a large cohort of patients, and a spectral analysis of each channel revealed consistent sites in sensorimotor cortex whose power changed specifically for each movement~\cite{Miller:2007}.

We performed a related analysis based on the DMD spectrum and derived motor maps of tongue and hand movements that are closely consistent with the previously described results~\cite{Miller:2007}. 
DMD spectra were computed for each epoch of the motor screen, and DMD modes within a low frequency band (8--32 Hz) and a high frequency band (76--100 Hz) were averaged to derive mean modes in each band.
Next the baseline low frequency and high frequency mean modes were subtracted from the movement epochs to produce the frequency-dependent sensorimotor maps in \textbf{Figure~\ref{fig:motor_validation}}.

In the low frequency band, there was a generalized decrease in correlation across sensorimotor cortex in both the tongue and hand movement epochs. 
These decreases are visualized by decreases in relative DMD mode magnitude across the most posterior columns of the electrode array (\textbf{Figure~\ref{fig:motor_validation}a, b}).
In the high frequency band, we found relatively local groups of electrodes within sensorimotor cortex where DMD mode amplitude increased selectively for tongue and hand movement.
The foci for hand movement were more dorsal than the tongue movement sites.

\begin{figure*}[t]
\centering
\includegraphics[width=1\textwidth]{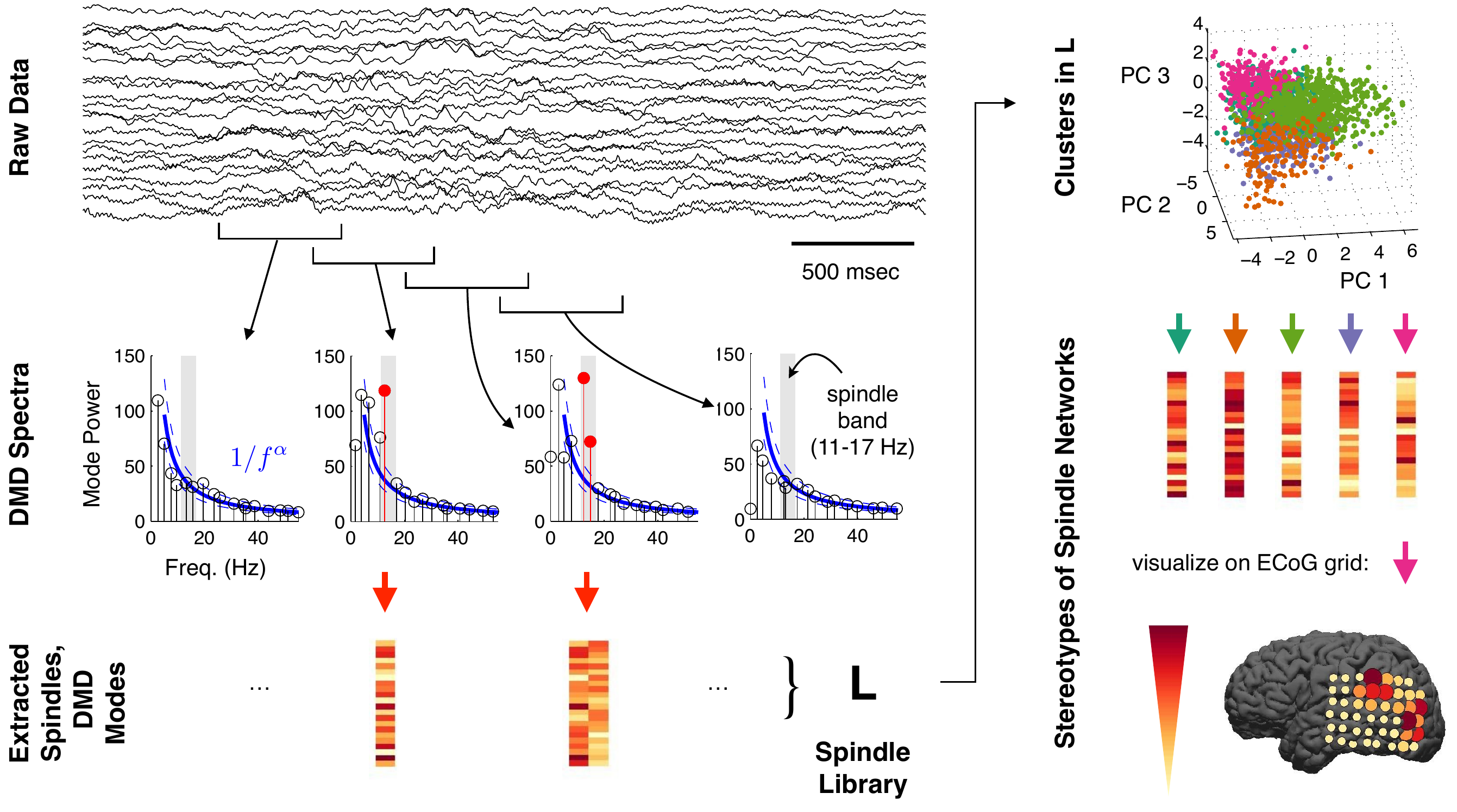}
\caption{A summary of how spindle networks were extracted based on the DMD spectra of windowed ECoG data. DMD spectra were computed for windows of raw data, and DMD modes in the spindle band (11--17 Hz, gray) whose power significantly exceeded the expected $1/f^\alpha$ distribution (blue lines) were extracted and gathered into $L$. This library $L$ was then clustered using a gaussian mixtures model to discover distinct spindle networks, and the centroids of these clusters were stereotypes of spindle networks that could be visualized on the ECoG grid. Each spindle network represented correlated groups of electrodes with activity in the spindle band.}
\label{fig:spindle_methods}
\end{figure*}

\subsection{Sleep spindle network extraction}\label{ssec:spindlenetwork}

In a novel application of DMD, we developed methods to extract and identify sleep spindle networks in human ECoG recordings.
Sleep spindles are distinctive, transient 12--15 Hz oscillations that are characteristic of non-rapid eye movement (NREM) sleep, and their presence is commonly used to classify sleep stages~\cite{DeGennaro:2003}.
Spindles have been the subject of scientific investigation since the early 1930's and their mechanisms of generation are now quite well understood~\cite{Steriade:1993}.
In brief, sleep spindles oscillations are generated in the thalamus and their electrographic signature arises from thalamacortical connections.
Even so, the role these transient oscillatory events play in brain function remains unclear.
A line of evidence suggests that sleep spindles facilitate the consolidation of recently acquired memories~\cite{Gais:2002, Clemens:2005, Eschenko:2006, Johnson:2012}.
This hypothesis is supported by recent work demonstrating that sleep spindles can be locally, rather than globally, synchronous events~\cite{Johnson:2012, Nir:2011}.

Historically, sleep spindles have been scored by experts on single channels of EEG.
Spindles vary in amplitude, duration, central frequency, and often concur with other regularly observed sleep features.
Automated detection algorithms typically rely on band-pass filtering the signal followed by an amplitude threshold on some moving average window (for instance~\cite{Schimicek:1994,Ray:2010}).
Recently, a number of these algorithms were evaluated against experts and crowd-sourced spindle detectors~\cite{Warby:2014}.
It is important to point out that all of these approaches only address spindle detection on single electrodes, and reliable identification of networks of electrodes showing synchronous spindle activity has remained a challenge.
The structure of sleep spindle networks may reveal insights about thalamacortical connections and other neural circuitry, which motivates development of more sophisticated methods for their detection and characterization.

Our approach to spindle network extraction is based on the DMD spectra of windowed ECoG data (\textbf{Figure~\ref{fig:spindle_methods}}). 
DMD modes corresponding to larger than expected power in the spindle band (11--17~Hz) were collected in a library, whose elements were then clustered to determine stereotypes of spindle networks in the recording. 
Each spindle network stereotype represents strength of spatial correlations between electrodes and may be visualized on a grid of electrode locations.

\begin{figure*}[t]
\centering
\includegraphics[width=1\textwidth]{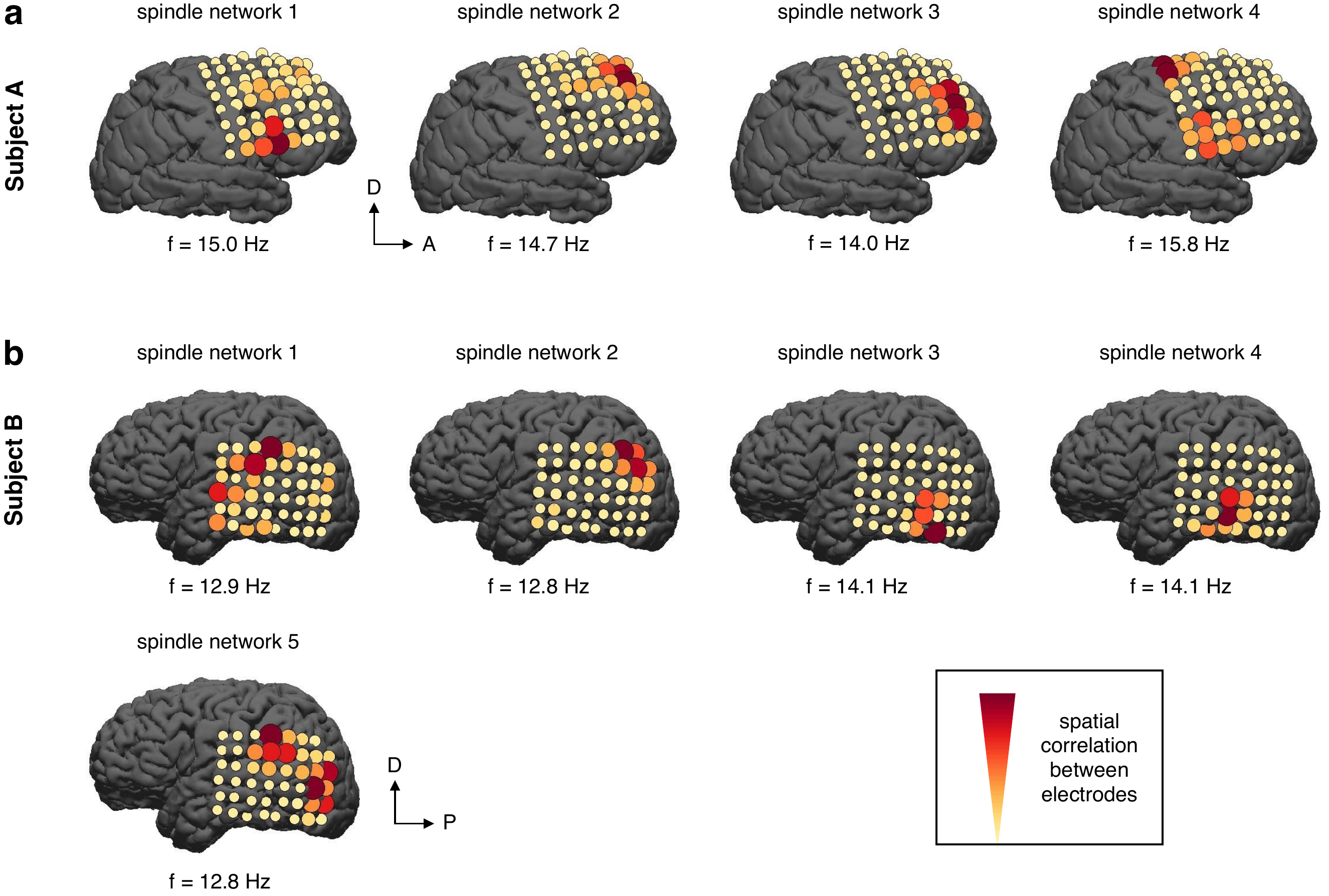}
\caption{Visualization of spindle network stereotypes for two subjects, computed from ECoG recordings during non-REM sleep. The spindle networks are shown on the ECoG recording grid, where both the size and color of dots at each electrode represent the strength of the spatial correlation between electrodes (see color bar).}
\label{fig:spindlestereotypes}
\end{figure*}

We applied our spindle network extraction algorithm to ECoG recorded during non-REM sleep from sub-dural electrodes implanted over the temporal lobe of two human subjects.
We found distinct stereotypes of spindle networks, whose spatial distributions are visualized for in \textbf{Figure~\ref{fig:spindlestereotypes}}.
Subject A had 5 distinct spindle networks over a 48-channel array; subject B had 4 distinct spindle networks over a 64-channel array. 
In both subjects, we found both relatively local and more spatially distributed spindle networks. 

In addition to separation of spindle activity to distinct locations on the cortex, spindle networks were also distinct from each other by their characteristic frequencies of oscillation.
These frequencies were not used in the clustering.

\section{Discussion}\label{sec:discussion}

In this work, we presented the novel adaption of a modal decomposition technique known as dynamic mode decomposition (DMD) to analyze and visualize large-scale neural recordings.
Such neural recordings are often characterized by coherent patterns of activity in space (across many channels) at multiple temporal frequencies.
We showed how DMD can be used to extract these coherent patterns by decomposing the data into a low-dimensional representation in both space and time.
This extraction was validated on a simple motor task where selective and separable regions of sensorimotor cortex were identified.

Next, we presented a novel method to detect and analyze sleep spindle networks from large-scale human ECoG recordings; automated approaches to characterize these stereotyped spindle networks had not been previously reported, and the high-dimensionality of the dataset makes manual scoring by experts an infeasible task.
We showed that our method reliably detected spindle networks, which are correlated groups of electrodes that showed significantly power in the spindle band (between 11 and 17 Hz).
Further, we found multiple stereotypes of spindle networks for two different subjects.
Some of these spindle networks were fairly localized on the cortex, but some revealed correlations across physically non-adjacent parts of cortex.
These sleep spindle networks are discovered independently of an explicit behavior and may represent functionally connected cortical areas.

The automated identification of spindle networks and their classification into distinct stereotypes allows us to analyze the duration, frequency, and relative timing of potentially independent spindle events.
In addition, we are now able to compare their timing relative to other events common in sleep (e.g., K-complexes) and relative to pre-epileptic indicators.
Our method enables these analyses to be performed across many subjects, and its computation scales favorably with increasing numbers of channels in the electrode array recordings.

Dimensionality reduction is a useful concept in building meaningful models based on dynamics of neuronal networks because there exists low-dimensional structures in large-scale data. 
Ideas from compressive sensing further suggest that such models may be accessible even from limited data, when the system is known to be sufficiently sparse~\cite{Ganguli:2012}.
Extensions of DMD that exploit underlying data sparsity (including~\cite{Jovanovic:2014, Brunton:2013}) have potential to expand the usefulness of the framework to a larger class of incomplete measurements with improved robustness.

In contrast to other static modal decomposition techniques, DMD provides not only modes, but also a relatively low-dimensional, efficient linear model for how the most dynamically important modes evolve in time.
In other words, the DMD approximation of the dataset as shown in \textbf{Figure~\ref{fig:dmd}} and expressed by Equation~\eqref{eq:dmdrecon} is also a prediction of the system's dynamic trajectory.
Such dynamic models of high-dimensional dynamic data are \emph{equation-free}, in the sense that they are entirely data-driven and do not rely on a set of governing equations.
Model predictions do not typically hold for long term; however, even short-term DMD predictions can inform design of feedback control in complex, nonlinear systems~\cite{Potter:2014}.

We propose that DMD is generally useful for the analysis, understanding, and visualization of large-scale neural recordings.
In particular, we are exploring DMD-based methods to characterize network-level responses across multiple brain areas to complex stimulus inputs.
We envision this spatial-temporal modal decomposition may be applied to a variety of modes of measurements, including ECoG, MEG, functional MRI, calcium imaging, LFP, and spike rates of a large collection of single units.

\section{Methods}

\subsection{Collection and preprocessing of ECoG recordings}

\subsubsection{Subjects}
All subjects were patients undergoing long-term electrocorticography (ECoG) monitoring in preparation for surgical treatment of intractable epilepsy.  Data were collected from two subjects (female subjects A and B, ages 29 and 11 respectively) with subdural platinum electrode arrays (Ad-Tech, Racine, WI).  Electrodes had a 2.3mm exposed surface diameter and were spaced at 1 cm.  Subjects were patients at the University of Washington and gave informed consent according to the protocol approved by the Institutional Review Board of that institution. 

\subsubsection{Recording}

The ECoG signals were recorded by the XLTEK (Natus Medical Incorporated, San Carlos, California) clinical monitoring system at a sampling rate of 500 or 2000Hz.  The standard system parameters impose a high-pass filter at about 0.1Hz. 
For the motor mapping, recordings were high pass filtered above 6 Hz and down sampled to 100 samples/sec; the total length of the motor screen recordings were approximately 4 minutes.
For spindle network extraction, recordings were bandpass filtered between 6 and 80 Hz, and down sampled at 200 samples/sec; the total length of recording of a sleep epoch from each subject was approximately 20 to 40 minutes.
Subject A had an electrode array with 64 channels; Subject B had an electrode array with 48 channels.

\subsubsection{Motor screen task}

The simple motor screen task was performed as described previously in~\cite{Miller:2007}.
Briefly, the patient was instructed either not move (baseline), move their tongue, or move the hand contralateral to the implanted electrode array.
Twenty (20) trials of each movement were repeated in pseudo-random order with instructions to not move in between; each instruction lasted three (3) seconds.
The data in this analysis was from Subject A.

\subsubsection{Sleep identification}

Non-REM sleep epochs were identified by eye using the increased power in the delta band (1--6Hz) and verified by the presence of K-complexes and spindles.  For every sleep epoch, all electrode traces were normalized by z-scoring with respect to the amplitude in the 5--50Hz range.  This frequency range was chosen to eliminate the variable amplitude effects of K-complexes, which have a maximum amplitude in medial-frontal regions 28.

\subsection{Sensorimotor mapping}

To map cortical areas whose activity were modified during movement of the tongue and hands in a frequency-dependent way, we divided the electrode array recording into baseline (no instructed movement), tongue movement, and hand movement trials.
The middle 2 seconds of each trial (between 500 and 2500 msec from the beginning of the instruction) where used in the motor mapping analysis.
In each trial, the 2 second recording window over all channels was decomposed by DMD; the time series data were augmented by stacking 10 times as described in the main text to increase the rank of the data matrix, and the resultant basis was truncated using $r=200$ so that the result spectrum contained only 200 modes.
The DMD spectrum was examined to extract spatial modes in two frequency ranges, a low frequency band (8--32 Hz) and a high frequency band (75--100 Hz). 
All the absolute value of DMD modes in each frequency range were averaged to extract the mean mode magnitude.
Next, for both the low and high frequency bands, we compared the mean mode across all trials for hand and tongue movements by subtraction of the mean baseline modes.
The relative changes in these mean mode magnitudes were reshaped and visualized at the electrodes' implanted locations in \textbf{Figure~3}.

\subsection{Spindle network extraction}
To detect the presence of spindle networks, we first separated the ECoG recordings into 300 msec windows in time, decomposed the windowed data using DMD, then used the DMD spectrum to identify modes that represent a significant increase in the 11--17 Hz frequency band characteristic of spindle activity. We used overlapping windows, sliding 100 msec between consecutive windows.

The DMD power spectrum allows us to formulate a principled criterion to identify the presence of power in the spindle band.
We identified power between 11 and 17 Hz, above what would have been expected from the power spectral density of ECoG recordings.
Next, DMD modes corresponding to excess power in the spindle band were collected over the entire recording epoch and stereotypes of spindle networks were discovered by a gaussian mixtures model.

\subsubsection{Spindle detection criteria}
It has been well described that the power spectral density of recordings from the brain follows a $1/f^\alpha$ distribution, where $\alpha$ is a positive scalar~\cite{Bedard:2006, Miller:2009}.
We used this observation to fit the $\alpha$ of the DMD spectrum, excluding power at frequencies below 18 Hz and above 57 Hz. 
This $1/f^\alpha$ distribution was fit over the cumulative DMD spectra of each epoch of a subject's recording (usually about an hour) using robust regression~\cite{Holland:1977}, which rejected outliers due to sporadic electrical noise in the recordings.
Windows including presumptive epileptiform activity were rejected base on a generalized elevation in power over all frequencies.
For each window, a DMD mode $\phi$ was considered part of a spindle network if and only if its power significantly exceeded the upper 99\% confidence interval of the  $1/f^\alpha$ fit.
Further, to reduce the incidence of spurious identification of spindle networks, significant power in the spindle band must be detected for three consecutive overlapping windows.

\subsubsection{Clustering spindle networks}
From every window of data, DMD modes that passed the spindle detection criteria were gathered to form a library of spindle networks $\bL$, and elements of this library were clustered into distinct types of spindle networks.
Specifically, we constructed a library of DMD modes $\bL$,
\begin{align*}
\bL &= \begin{bmatrix}
\vline & \vline & & \vline \\
\phi_1 & \phi_2 & \cdots & \phi_{N}\\
\vline & \vline & & \vline
\end{bmatrix}.
\end{align*}
For purposes of clustering, we considered only the absolute value of DMD modes; each mode was normalized to unit length.
Clusters were determined in $r$-dimensional principal components space, using the projections of each column of $|\bL|$ onto the first $r$ principal components of $|\bL|$:
\begin{align}
|\bL| &= \bU_L \bSigma_L \bV_L^*, \\
\ba &= \bU_r^T \bL,
\end{align}
where $\bU_r^T$ is the transpose of the first $r$ columns of $\bU_L$.

Next we used a gaussian mixtures model to group columns of $\ba$ as points in $\mathbb{R}^r$ into $k$ clusters~\cite{Murphy:ML}.
Since this is an unsupervised approach, we selected the appropriate value of $k$ in the model using the Bayes Information Criterion (BIC), a statistical metric to guide comparison of models with different numbers of parameters that punishes overfitting.
Briefly, a model with a smaller BIC is more parsimonious with the available data than a model with a larger BIC.
We evaluated descriptions of $|\bL|$ using a gaussian mixtures model in $r$ dimensions using $k$ and systematically varied $r$ from 3--15 and varied $k$ from 2--10.
The number of stereotyped spindle networks $k$ was chosen to minimize BIC of this family of gaussian mixtures model.
For the two subjects we tested, the median $k$ for which BIC was minimized was 4 and 5 clusters for subjects A and B, respectively. 
We chose to fit the gaussian mixtures model with $r=5$ reduced dimensions, because increasing $r$ beyond $5$ did not significantly change the assignment of clusters.

\section*{Acknowledgements}
We thank S. L. Brunton and J. L. Proctor for comments on the manuscript,  J. P. Dyhr and T. L. Daniel for helpful discussions, and T. Blakely for collecting the dataset used for the sensorimotor mapping. 
This research was supported by National Institutes of Health Grant R01 NS065186-01 and award number EEC-1028725 from the National Science Foundation.


\section*{Author Contributions}
B.W.B. and L.A.J. conceived of the project; L.A.J. collected the sleep data. B.W.B. designed the analyses and implemented the algorithms, with contributions from L.A.J. J.G.O. and J.N.K. supervised the project. B.W.B. wrote the manuscript, with input from all authors.

\bibliographystyle{ieeetr} 
\footnotesize{
\bibliography{dmd_ecog}
}

\end{document}